\documentstyle{article}

\author{M.I. Piso, N. Ionescu-Pallas, S. Onofrei\\
Gravitational Researches Laboratory\
71111 Bucharest, Romania}
\title{LINEAR  BIMETRIC  GRAVITATION  THEORY}

\begin{document}
\maketitle
\begin{abstract}

A general bimetric theory of gravitation is described as a linear
in the second approximation. This is allowed due to the small
experimental significance of the higher order
terms. Solar System tests are satisfied. The theory allows black holes, which
are physical singularities. The predicted black hole radius is equal to the one
resulting from Einstein's theory, if the compatibility between the
gauge conditions and the existence of gravitational waves is required
(the Rosen-Fock metric). The quadrupolar gravitational radiation
formula is regained.

\bigskip\

\end{abstract}
\
\section{Introduction}

Bimetric theories are considered as a main result of Nathan Rosen's work in
the early 40's \cite{Rosen1}. The direct interest in this way of space-time
and gravity interpretation is related to the possibility to describe the
gravitational interaction as a physical field, acting directly onto the
standard flat space-time as a result of some ''geometric'' properties of a
locally attached general relativistic manifold. The resulting field
description of gravity may suggest standard field theory methods in order to
quantize the gravitational field. However, the efforts in this direction are
discouraged by some non convenient features of the Rosen's theory: $i$)
black holes do not exist and $ii$) the interpretation of the observational
gravitational radiation emission results from the PSR 1913+16 pulsar fails
in comparison with general relativity.

The present paper deals with some opinions on the physical possibilities in
order to make from bimetric theories good gravitational field theories, by
removing the incompatibilities with general relativity and observational data.

\section{Bimetric theory}

In the sequel, we shall try to give a physical description of the basic
bimetric concepts. To conciliate both general relativity and the standard
Minkowski description of quantum field theories, two four-dimensional
''Universes'' are simultaneously considered: one of them is flat and the
other curved \cite{Rosen1}. The flat Universe is the ''real'' one, in which
experiments and observations are performed. The curved Universe is
considered as a necessary geometric construction in order to satisfy the
inertial and gravitational mass equivalence principle \cite{Rosen2}\cite
{Papapet}\cite{Kohler}.

For both ''Universes'', we adopt the same arbitrar coordinate system. To
evidence the separation between the ''Universes'', we denotie by the sign $R$
the quantities defined on the curved Universe. The metrics in the two
''Universes'' are:

$$
(dS)^2=g_{\alpha \beta }dx^\alpha dx^\beta \;\;,\;\;\;\;\;(dS_R)^2=\gamma
_{\alpha \beta }dx^\alpha dx^\beta
$$
\begin{equation}
\label{1}(\alpha ,\beta )=(0,1,2,3)\;\;,\;\;\;\;dS>0\;\;,\;\;\;\;dS_R>0
\end{equation}
where $g_{\alpha \beta }$ and $\gamma _{\alpha \beta }$ are the metric
tensors for the considered cases. Contravariant equivalent metric tensors
are defined by means of the relations:
\begin{equation}
\label{2}g_{\alpha \lambda }h^{\lambda \beta }=\delta _\alpha ^{\;\;\beta
}\;\;,\;\;\;\;\gamma _{\alpha \lambda }\chi ^{\lambda \beta }=\delta _\alpha
^{\;\;\beta }
\end{equation}
where $\delta _\alpha ^{\;\;\beta }$ are the Kronecker symbols.

Since bimetric concepts, the physical ''meaning'' of the curved Universe is
established by giving a relation between the contravariant metric tensor $%
\chi ^{\alpha \beta }$ and a necessary tensor potential $\Psi ^{\alpha \beta
}$ of the gravitational field, which lies on the flat Universe \cite{Rosen2}
In this paper, we shall restrict to the second order terms of the general
relation, due to the small experimental significance of the higher order
terms:

\begin{equation}
\label{3}\chi ^{\alpha \beta }=h^{\alpha \beta }-\epsilon \Psi ^{\alpha
\beta }+\epsilon ^2(b_1\Psi \Psi ^{\alpha \beta }+b_2\Psi ^{\alpha \lambda
}\Psi _\lambda ^{\;\;\beta }+b_3\Psi ^2h^{\alpha \beta }+b_4\Psi _{\mu \nu
}\Psi ^{\mu \nu }h^{\alpha \beta })
\end{equation}
where $\Psi =\Psi ^{\alpha \beta }\Psi _{\alpha \beta }$; $b_j$ ($j=1,2,3,4$%
) are dimensionalless quantities to be determined in the sequel, and $%
\epsilon $ is a parameter whose physical dimensions are the inverse of the
ones of the gravitational potential $\Psi $. Outside the sources, it seems
obvious from \ref{3} that $\epsilon \Psi \ll 1$. It is convenient to adopt
for $\epsilon $ the expression:
\begin{equation}
\label{4}\epsilon =\frac{GM_0}{c^2}
\end{equation}
which holds for a finite physical system with rest mass $M_0$; $G$ is the
Newtonian gravity constant and $c$ - the speed of light.

To resume, geometry on the curved space-time is determined by the metric
tensor $\chi ^{\alpha \beta }$, the curvature being produced by the flat
space gravitational tensor potential $\Psi ^{\alpha \beta }$ as seen from
\ref{3}.

A basic principle of the bimetric theories is the existence of a variational
principle of the action $A$:
\begin{equation}
\label{5}A=\frac 1c\int L\sqrt{-g}(d^4x)
\end{equation}

where the Lagrangeian density $L$ is composed of a sources part, formulated
in the curved Universe, and a field part, formulated in the flat Universe
\cite{Rosen3}\cite{Rosen4}:
\begin{equation}
\label{6}L=S\left[ \rho c^2+\left( \rho H-p\right) \right] _R-\frac{c^4}{%
64\pi G}\epsilon ^2\left( \Psi _{\alpha \beta \mid \lambda }\Psi ^{\alpha
\beta \mid \lambda }-\frac 12\Psi _{\mid \lambda }\Psi ^{\mid \lambda
}\right)
\end{equation}
The quantities $S$ and $H$ are as follows:
\begin{equation}
\label{7}S=\frac{\sqrt{-\gamma }}{\sqrt{-g}}\;\;,\;\;\;\;H=\int\limits_0^{p(%
\rho )}\frac{dp}{\rho (p)}\;\;,\;\;\;\;\gamma =Det\left\| \gamma _{\alpha
\beta }\right\| \;\;,\;\;\;\;g=Det\left\| g_{\alpha \beta }\right\|
\end{equation}
\ By means of \ref{6} and \ref{7}, the physical system to be studies is
composed of a perfect fluid, described by the invariant mass density $\rho $%
, the invariant pressure $p$ and the tensor potential of the system's proper
gravity $\Psi ^{\alpha \beta }$.

Performing the variation of the action \ref{5} with respect to $\Psi
^{\alpha \beta }$ and vanishing it, we get the field equations \cite{NJP1}:
\begin{equation}
\label{8}\frac 12\epsilon \Box \xi _{\alpha \beta }=-\frac{8\pi G}{c^4}%
\sigma _{\alpha \beta }
\end{equation}
where
\begin{equation}
\label{9}\xi _{\alpha \beta }\equiv \Psi _{\alpha \beta }-\frac 12g_{\alpha
\beta }\Psi \;\;,\;\;\;\;\sigma _{\alpha \beta }\equiv -\frac 1\epsilon
ST_{\mu \nu }\frac{\partial \chi ^{\mu \nu }}{\partial \Psi ^{\alpha \beta }}
\end{equation}
and $T_{\mu \nu }$ is the energy tensor defined in the curved Universe:
\begin{equation}
\label{10}T_{\alpha \beta }=\left[ \left( c^2+H\right) \rho u_\alpha u_\beta
-p\gamma _{\alpha \beta }\right] _R
\end{equation}

Explicitly from \ref{8} and \ref{9}, $\sigma _{\alpha \beta }$ - the tensor
of the gravitational sources - has the form:
\begin{equation}
\label{11}
\begin{array}{c}
\sigma _{\alpha \beta }=S\{T_{\alpha \beta }-\epsilon b_1\left( \Psi
T_{\alpha \beta }+\Psi ^{\mu \nu }T_{\mu \nu }g_{\alpha \beta }\right)
-\epsilon b_2\left( \Psi _\beta ^{\;\;\lambda }T_{\lambda \alpha }+\Psi
_\alpha ^{\;\;\lambda }T_{\lambda \beta }\right) \\
-2\epsilon b_3\Psi \left( h^{\mu \nu }T_{\mu \nu }\right) g_{\alpha \beta
}-2\epsilon b_4\Psi _{\alpha \beta }\left( h^{\mu \nu }T_{\mu \nu }\right)
\}
\end{array}
\end{equation}

One considers the next two approximations: the first one, derived from the
first terms expansion of \ref{10}, is
\begin{equation}
\label{12}ST_{\alpha \beta }\approx \left[ \left( c^2+H\right) \rho u_\alpha
u_\beta -pg_{\alpha \beta }\right] +\epsilon c^2\rho \left( u_\alpha
u^\lambda \Psi _{\lambda \beta }+u_\beta u^\lambda \Psi _{\lambda \alpha
}-\frac 12u_\alpha u_\beta \Psi _{\mu \nu }u^\mu u^\nu \right)
\end{equation}
and the second stands only for the terms proportional to $\epsilon $ in \ref
{11}
\begin{equation}
\label{13}ST_{\alpha \beta }\approx \rho c^2u_\alpha u_\beta
\end{equation}
In this way, we get for the gravitational field sources tensor an
approximative form, which is covariant in the flat Universe:

\begin{equation}
\label{14}
\begin{array}{c}
\sigma _{\alpha \beta }\approx \left[ \left( c^2+H\right) \rho u_\alpha
u_\beta -pg_{\alpha \beta }\right] +\epsilon \rho c^2\{-b_1\left[ \Psi
u_\alpha u_\beta +\left( u_\mu u_\nu \Psi ^{\mu \nu }\right) g_{\alpha \beta
}\right] \\
+\left( 1-b_2\right) \left[ u_\alpha u^\lambda \Psi _{\lambda \beta
}+u_\beta u^\lambda \Psi _{\lambda \alpha }\right] -2b_3\Psi g_{\alpha \beta
}-2b_4\Psi _{\alpha \beta }-\frac 12\left( u_\mu u_\nu \Psi ^{\mu \nu
}\right) u_\alpha u_\beta \}
\end{array}
\end{equation}
Further, considering a covariant approximation for the potential and
introducing a new denotation instead:
\begin{equation}
\label{15}\Psi _{\alpha \beta }\approx \Psi \left( \frac 12g_{\alpha \beta
}-u_\alpha u_\beta \right) \;,\;\;\;\;\;\;\Psi =\frac 4{\epsilon c^2}\Phi
\end{equation}
we get for \ref{14}
\begin{equation}
\label{16}\sigma _{\alpha \beta }\approx \left[ \left( c^2+H\right) \rho
u_\alpha u_\beta -pg_{\alpha \beta }\right] +4\rho \Phi \{\left[ \frac
12b_1-2b_3-b_4\right] g_{\alpha \beta }+\left[ 2b_4+b_2-b_1-\frac 34\right]
u_{\alpha \beta }\}
\end{equation}
In order to satisfy the equivalence principle for the one body problem, in
the case
\begin{equation}
\label{17}g_{\alpha \beta }\longrightarrow 2\delta _{0\alpha }\delta
_{0\beta }-\delta _{\alpha \beta }\;,\;\;\;\;u_\alpha \longrightarrow \delta
_{0\alpha }
\end{equation}
the following expression should be fulfilled
\begin{equation}
\label{18}\sigma _{00}\approx c^2\rho +\left( \rho H-p\right) -\frac 12\rho
\Phi \;\;.
\end{equation}
{}From \ref{16}, \ref{17} and \ref{18} we get the first relation between the
coeficients:

\begin{equation}
\label{19}4b_4-8b_3+4b_2-2b_1=\frac 52
\end{equation}
In the static approximation, $\gamma _{\alpha \beta }$ becomes
\begin{equation}
\label{20}
\begin{array}{c}
\gamma _{\alpha \beta }=\left( 2\delta _{0\alpha }\delta _{0\beta }-\delta
_{\alpha \beta }\right) -\frac 12\epsilon \Psi \delta _{\alpha \beta
}+\epsilon ^2\Psi ^2\{\left( \frac 12-\frac 12b_2-2b_3-2b_4\right) \delta
_{0\alpha }\delta _{0\beta }- \\
\left( \frac 14-\frac 12b_1-\frac 14b_2-b_3-b_4\right) \delta _{\alpha \beta
}\}+O(\epsilon ^3\Psi ^3)
\end{array}
\end{equation}
To satisfy the equivalence principle in the two-body problem, one should
have \cite{NJP2}
\begin{equation}
\label{21}\gamma _{00}=1-\frac 12\epsilon \Psi +\frac 18\epsilon ^2\Psi ^2
\end{equation}
{}From \ref{20} and \ref{21} we get the second condition for the coefficients
$%
b_i$%
\begin{equation}
\label{22}4b_4+4b_3+b_2-2b_1=\frac 12
\end{equation}
and, using also \ref{19}, we get the solutions:
\begin{equation}
\label{23}b_1=\frac 1{12}+\left( 4b_3+2b_4\right) \;\;,\;\;\;\;b_2=\left(
\frac 23+4b_3\right) \;\;.
\end{equation}

In the static case, $\chi ^{\alpha \beta }$ is a diagonal tensor which may
be exactly inversed \ref{2} in order to get $\gamma ^{\alpha \beta }$ and
get for the metric \ref{1} the Schwarzschild type form
\begin{equation}
\label{24}\left( dS_R\right) ^2=\frac{\left( cdt\right) ^2}{1+\frac
12\epsilon \Psi +\frac 18\epsilon ^2\Psi ^2}-\frac{\left( dx\right)
^2+\left( dy\right) ^2+\left( dz\right) ^2}{1-\frac 12\epsilon \Psi
+\epsilon ^2\Psi ^2\left( \frac 18+2\eta \right) }
\end{equation}
$$
\eta \equiv b_4+2b_3+\frac 1{24}
$$

Performing the variation of the action with respect to the coordinates of
the fluid particle we get that, outside the field sources, in the curved
Universe metric, the probe particle movement is geodetic. On the other hand,
if we take in \ref{24} the $\epsilon $ expansions in the second order for $%
\gamma _{00}$ and $\gamma _{jk}$ ($j,k=1,2,3$) we get an expression which,
in the same approximation, is to coincide to the homologous expression in
standard general relativity \cite{Weinberg}. This correspondence implies
that the Solar system general relativistic tests are satisfied.

If we take $\eta =0$, i.e. from \ref{24}
\begin{equation}
\label{25}\left( b_4+2b_3+\frac 1{24}=0\right)
\end{equation}
the metric \ref{24} may be considered as an approximation of the Rosen's
metric \cite{Rosen3} written as following:
\begin{equation}
\label{26}(dS_R)^2=\frac{(cdt)^2}{\exp (\frac 12\epsilon \Psi )}-\frac{%
(dx)^2+(dy)^2+(dz)^2}{\exp (\frac 12\epsilon \Psi )}
\end{equation}
However, for $r\longrightarrow 0$, $\Psi \propto \frac 1r$ in the case of a
pointlike source, and the metric \ref{26} manifests exponentially divergent
in the origin, different from the behaviour of the metric \ref{24}. If we
take in \ref{24} the values of $\eta >-\frac 1{32}$, then the denominators
of the metric coefficients do not vanish for real values of $r$, i.e. no
Schwarzschild-like black-holes exist. But, it is to emphasize that the
resultant metric \ref{24} does not exclude the existence of black holes. It
is obvious that, for
\begin{equation}
\label{27}\eta =-\frac 1{32},\;\;i.e.\;\;\;\left( b_4+2b_3+\frac
7{96}=0\right)
\end{equation}
the denominator of the spacelike part of the metric \ref{24} becomes $\left(
1-\frac 14\epsilon \Psi \right) ^2$ and the resulting radius of the
singularity is:
\begin{equation}
\label{28}r=\epsilon
\end{equation}

The important conclusion is that bimetric theories may provide a more
general formalism for the description of gravity. It is interesting to
remark that Rosen himself believed that bimetrism is incompatible with black
holes \cite{Rosen4}. Is is to be mentioned that the predicted black hole
radius \ref{28} is equal with the one resulting from Einstein's theory, if
the compatibility between the gauge conditions and the existence of
gravitational waves is required (the Rosen-Fock metric) \cite{Fock}.
\

\
\newpage
\

\section{Appendix}

To study the gravitational radiation, we start from the following
approximations:

\begin{equation}
\label{29}\epsilon \xi ^{\alpha \beta }\approx \frac{4G}{c^4}\int \frac{%
\left[ \tau ^{\alpha \beta }\right] }Rdxdydz\;\;,\;\;\;\;\xi
_{\;\;\;\;,\beta }^{\alpha \beta }\approx 0\,\,,\,\,\,\,\tau ^{\alpha \beta
}\approx \rho c^2u^\alpha u^\beta
\end{equation}
Performing a standard procedure of approximation \cite{Landau} we reach to
the following formula:
\begin{equation}
\label{30}\epsilon \xi ^{jk}\approx \frac{2G}{c^4R_0}\frac{d^2}{dt^2}\int
\rho x^ix^jdxdydz
\end{equation}

With the standard definition for the mass quadrupole momentum:
\begin{equation}
\label{31}D^{jk}=\int \rho \left( 3x^jx^k-\delta ^{jk}\delta
_{lm}x^lx^m\right) dxdydz
\end{equation}
we get the known relation:
\begin{equation}
\label{32}\epsilon \left( 3\Psi ^{jk}-\delta ^{jk}\delta _{lm}\Psi
^{lm}\right) =\frac{2G}{c^4R_0}\stackrel{..}{D}^{jk}
\end{equation}
In the following, we introduce the denotations:
\begin{equation}
\label{33a}\alpha =\frac{2G}{c^4R_0}\stackrel{..}{D}^{j11},\;\;\;\;\beta =
\frac{2G}{c^4R_0}\stackrel{..}{D}^{j22},\;\;\;\gamma =\frac{2G}{c^4R_0}%
\stackrel{..}{D}^{j33}
\end{equation}
take into account of the dyadic character of the $D^{jk}$ tensor, i.e.
\begin{equation}
\label{33b}\alpha =\beta =\gamma
\end{equation}
and the approximate
\begin{equation}
\label{34}\Psi ^{\alpha \beta }=A^{\alpha \beta }\sin \omega \left( t-\frac
xc\right)
\end{equation}
\begin{equation}
\label{36}\phi ^\alpha \equiv q^\alpha \left[ 1-\cos \omega \left( t-\frac
xc\right) \right]
\end{equation}
\
\begin{equation}
\label{37a}\Psi _{\mu \nu }^{\prime }=\Psi _{\mu \nu }+\left( g_{\mu l}{\bf D%
}_\nu +g_{\nu \lambda }{\bf D}_\mu \right) \phi ^\lambda
\end{equation}
(approximate gauge transformation for avoiding longitudinal waves)

\begin{equation}
\label{44}c\theta _0^{\;\;1}=\frac{c^3}{32\pi G}\epsilon ^2\frac{\partial
\Psi _{\alpha \beta }^{\prime }}{\partial t}\frac{\partial \Psi ^{\prime
\alpha \beta }}{\partial t}
\end{equation}
(supplemented by the condition $\Psi ^{\prime }=0$ accomplished as a result
of the gauge transformation)
\begin{equation}
\label{45}P=-c\oint \theta _0^{\;\;1}R_0^2d\Omega =-\frac G{45c^5}\stackrel{%
...}{D}_{jk}\stackrel{...}{D}^{jk}
\end{equation}

\end{document}